\documentclass{amsart}
\usepackage{amssymb,graphics}
\usepackage{epstopdf}
\newtheorem{thm}{Theorem}[section]

\newtheorem{corollary}[thm]{Corollary}

\newcommand{\dontprint}[1]{\relax}
\newcommand{\bx}[1]%{{{\fbox{$\scriptstyle{#1}$}}}}
{\begin{picture}(8,8)\put(0.5,-1.0){\framebox(7,7){$\scriptstyle{#1}$}}\end{picture}}
\title[Baker-Akhiezer function as iterated residue and Selberg-type integral]
{Baker-Akhiezer function as iterated residue and Selberg-type integral}
\author{Giovanni Felder}
\address{Department of mathematics, ETH Zurich,
8092 Zurich, Switzerland}
\email{giovanni.felder@math.ethz.ch}
\author{Alexander P. Veselov}
\address{Department of Mathematical Sciences,
Loughborough University, Loughborough LE11 3TU, UK  and Landau
Institute for Theoretical Physics, Moscow, Russia}
\email{A.P.Veselov@lboro.ac.uk}
\begin{document}
\maketitle

\begin{abstract}
A simple integral formula as an iterated residue is presented for
the Baker-Akhiezer function related to $A_n$ type root system both
in the rational and trigonometric cases. We present also a formula
for the Baker-Akhiezer function as a Selberg-type integral and
generalise it to the deformed $A_{n,1}$-case. These formulas can be
interpreted as new cases of explicit evaluation of Selberg-type
integrals.
\end{abstract}

\section{Introduction}
 The notion of rational Baker-Akhiezer (BA) function related to a configuration of hyperplanes with multiplicities was introduced in \cite{CV, VSCh, CFV} as a multi-dimensional version of Krichever's axiomatic approach \cite{Kri}. Such function exists only for special configurations, in particular for all Coxeter configurations.  For the configuration of type $A_{n-1}$ with multiplicity $m$ it has the form
\begin{equation}
\label{form}
\Psi^{(n)}_m = \frac{P_m^{(n)}(x, \lambda)}{A(x)^m A(\lambda)^m} \exp(\lambda, x),
\end{equation}
 where
$A(x) = \prod_{i < j}^n (x_i-x_j), \, (\lambda,x) = \sum_{i=1}^n \lambda_i x_i$ and $P_m^{(n)}(x, \lambda)$ is a polynomial in both $x=(x_1, \dots, x_{n}) \in \mathbb C^n$ and $\lambda=(\lambda_1, \dots, \lambda_{n})\in \mathbb C^n$ with the leading term $A(x)^m A(\lambda)^m.$ It satisfies the Schr\"odinger equation
\begin{equation}
\label{Sch}
L^{(n)}_m\Psi^{(n)}_m= -(\lambda, \lambda) \Psi^{(n)}_m,
\end{equation}
 where $L_m$ is the $n$ particle {\em Calogero--Moser operator}
\begin{equation}
\label{CM}
L^{(n)}_m=-\Delta + \sum_{i < j}^n \frac{2m(m+1)}{(x_i-x_j)^2}.
\end{equation}
 The rational BA function $\Psi^{(n)}_m$  is determined uniquely  by these properties  and has a remarkable symmetry:
$$\Psi^{(n)}_m(z, \lambda) = \Psi ^{(n)}_m(\lambda, z)$$ (see \cite{CFV}). It plays an important role in the theories of commutative rings of differential operators, Huygens principle and quasi-invariants of Coxeter groups
\cite{ CV, CFV, FeiV, EG}.

Three ways for computing this function are known.  The first one due to Chalykh and one of the authors \cite{CV} is known only in the first non-trivial case $m=1$  and is a recursive formula in the number $n$ of variables.  The second one uses the iteration of the shift operator (increasing $m$ by one) by Heckman and Opdam, which can be effectively described using the Dunkl operators \cite {Heck}. The third one, based on the formula due to Berest \cite{B}, is the most general: it works for all locus configurations, see \cite{CFV}.

In this note we present a new formula for the BA function both in
rational and trigonometric cases as a simple iterated residue.  The
structure of the integrand is the same as in other integral formulas
known in the theory of Calogero--Moser system and related Jack
polynomials (see \cite{KK, Awata, OO, KMS}), but the
integration cycle is different and adapted only for the case of
integer multiplicities.

We present also another formula for the Baker-Akhiezer function as a
Selberg-type integral, which can be considered as analytic
continuation of the residue formula from $m$ to $-m-1$, which is an
obvious symmetry of the Calogero--Moser operator. The comparison of
these two formulas with other known forms of the BA function can be
interpreted as new explicit evaluation of the special Selberg-type
integrals, which are probably new. We present similar results also
in the deformed $A(n,1)$-case discovered in \cite{CFV0, CFV1}.

Our approach is based on a generalisation of the identity, which plays an
important role in the theory of Jack polynomials \cite{Stanley, Ma} and
in various versions used in \cite{KK, Awata, R, Lan}. In
particular, Langmann \cite{Lan} suggested a simple explanation
of this identity within the theory of Calogero--Moser models with
different types of particles \cite{KO, Sen}, which is very
convenient for us and allows to extend it for the general
$A(n,m)$ deformation \cite{SV1}.

\section{Rational BA function}

The following result can be considered as a version of the ``adding
particle" approach from \cite{CV}. Let us introduce for any two set
of variables $u_1,\dots, u_k$ and $v_1, \dots v_l$ the function
\begin{equation}
\label{A}
A(u,v)= \prod_{i=1}^k \prod_{j=1}^l (u_i-v_j).
\end{equation}
 We will also use the notation
$$\bar u = u_1 + \dots + u_k.$$

%%%
For fixed distinct $x_1,\dots,x_{k+1}\in\mathbb C$ let us choose the
cycle of integration $\sigma$ in $k$ integration variables $z_i$ as
a product of small circles $|z_i-x_i| = \epsilon$ around the first
$k$ points $x_i,  i= 1, \dots, k$ and denote by $dz$ the
differential form
$$dz = dz_1 \wedge dz_2 \wedge \dots \wedge dz_k.$$

\begin{thm}
\label{th1} The BA functions with $k$ and $k+1$ particles are
related by the following iterated residue formula
\begin{equation}
\label{formula1*}
 \oint_{\sigma}  \frac{A(x)^{m+1}A(z)^{m+1}} {A(z,x)^{m+1}} e^{\lambda_{k+1}(\bar x - \bar z)} \Psi^{(k)}_m(z, \lambda_1, \dots, \lambda_k) dz= C_1 \Psi^{(k+1)}_m(x, \lambda_1, \dots, \lambda_{k+1})
\end{equation}
where  $$C_1=C_1(k,m, \lambda)= \frac{(2\pi i)^k}{(m!)^k} \prod_{i=1}^{k} (\lambda_i-\lambda_{k+1})^m.$$
\end{thm}

Iterating this procedure we come to the following formula for the BA
function. Note that for $n=1$ the Calogero--Moser operator
(\ref{CM}) becomes simply the second derivative and
$$\Psi^{(1)}_m(x, \lambda) = e^{\lambda_1 x_1}.$$
By adding one integration variable at each step we will have $\frac{n(n-1)}{2}$ integration variables, which we denote $t_{i,j}$ with $1 \leq i \leq j \leq n-1.$ It is convenient also to denote
$t_{i,n}=x_i, i=1, \dots, n.$
The integrand has the following form (cf. \cite{KK,Awata})
%\begin{equation}
%\label{omega}
$$
\omega_m=
 \prod_{i \leq j, l\leq  j+1}^{n} (t_{i,j}  - t_{l,j+1})^{-m-1}
\prod_{1 \leq i < l \leq j \leq n-1} (t_{i,j} - t_{l,j})^{2+2m} \prod_{l<j} e^{(\lambda_{j} - \lambda_{j+1}) t_{l,j}} dt_{l,j}.
$$
%\end{equation}
We assume that $x_i$ are distinct and choose the cycle of integration $\Sigma$ as the product of circles $|t_{k,j} - x_k| = \epsilon (n-j)$ with $\epsilon$ small enough.

\begin{corollary}
For any given positive integer $m$
the rational Baker-Akhiezer function (\ref{form}) can be given by the following iterated residue formula
\begin{equation}
\label{int}
\Psi^{(n)}_m(x, \lambda) =  \left(\frac{m!}{2\pi i}\right)^{\frac{n(n-1)}{2}} e^{\lambda_n\bar x} A(x)^{1+m}A(\lambda)^{-m} \oint_{\Sigma} \omega_m.
\end{equation}
\end{corollary}

To explain another integral representation of BA function note that
the Calogero--Moser operator (\ref{CM}) is invariant under the
change $$m \rightarrow -1-m.$$ This leads to the following formula
for BA function as a Selberg-type integral \cite{Var}.

Let us assume for convenience that $x_i, \, i=1, \dots, k+1$ have distinct imaginary parts and
$\lambda_i-\lambda_j$ have negative real parts for all $i<j, i,j=1,\dots, k+1.$
Choose the contour of integration $\gamma$ such that $z_i= x_i + \tau_i, \, i=1, \dots, k$ with real variables $\tau_i$, changing from 0 to $\infty.$

\begin{thm}
The rational BA functions for $k$ and $k+1$ particles are related by the following Selberg-type integral formula
\begin{equation}
\label{formula3}
\int_{\gamma}  \frac{A(z,x)^{m}} {A(x)^{m}A(z)^{m}} e^{\lambda_{k+1}(\bar x - \bar z)} \Psi^{(k)}_m(z, \lambda_1, \dots, \lambda_k) dz = C_2 \Psi^{(k+1)}_m(x, \lambda_1,\dots, \lambda_{k+1})),
\end{equation}
where
$$C_2=C_2(m, k, \lambda)= (m!)^k \prod_{i=1}^k (\lambda_{k+1}-\lambda_{i})^{-m-1}.$$
\end{thm}

Consider now in the same variables $t_{i,j}$  the form
 $$\alpha_m=\omega_{-m-1}=
  \prod_{i \leq j, l\leq  j+1}^{n} (t_{i,j}  - t_{l,j+1})^{m}
\prod_{1 \leq i < l \leq j \leq n-1} (t_{i,j} - t_{l,j})^{-2m} \prod_{l<j} e^{(\lambda_{j} - \lambda_{j+1}) t_{l,j}} dt_{l,j}$$
and choose the integration contour $\Gamma$ by assuming that $t_{i,j} = t_{i,j+1} + \tau_{i,j}$ with real variables $\tau_{i,j}, 1\leq i \leq j=1,\dots,n-1$ changing from zero to infinity.

\begin{corollary}
For any positive integer $m$
the rational Baker-Akhiezer function (\ref{form}) can be given by the following Selberg-type integral
\begin{equation}
\label{int3}
\Psi^{(n)}_m(x, \lambda) =  ((-1)^{m+1}m!)^{-\frac{n(n-1)}{2}} e^{\lambda_n \bar x} A(x)^{-m}A(\lambda)^{m+1} \int_{\Gamma} \alpha_{m}.
\end{equation}
\end{corollary}

\begin{figure}
 \begin{picture}(200,250)(0,-30)
 \put(18,0){$z_n$}
 \put(53,30){$z_{n-1}$}
 \put(25,68){$z_{n-2}$}
 \put(100,110){\dots}
 \put(100,140){\dots}
 \put(55,170){$z_1$}
 \scalebox{0.5}{\includegraphics{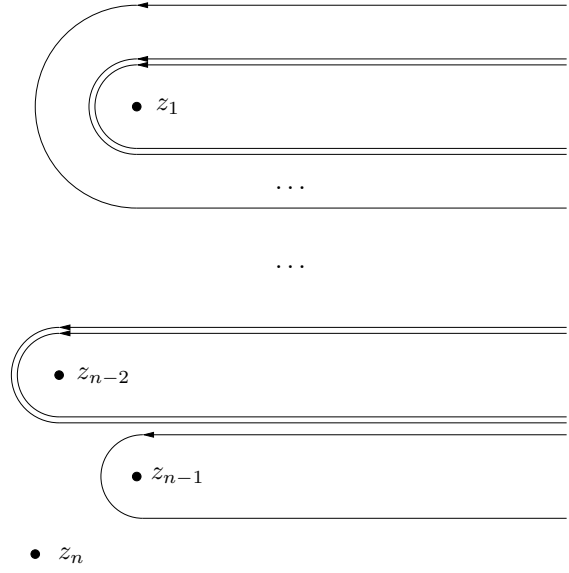}}
 \end{picture}
 \caption{The Pochhammer contour $\Pi$ in the rational case}\label{fig1}
 \end{figure}

These two different formulas are actually related by analytic continuation. To see this consider the same integral (\ref{int3}) but over Pochhammer contour $\Pi$:
$$I_P(m)= \int_{\Pi} \alpha_m$$
(see Figure \ref{fig1}).
 It converges for all $m \in \mathbb C$ and is related for positive real $m$ to the Selberg-type integral
 $$I(m)=\int_{\Gamma} \alpha_m$$ in a simple way:
 $$I_P(m) = (e^{2\pi i m} -1)^{\frac{n(n-1)}{2}} I(m).$$
Now replace $m!$ in (\ref{int3}) by $\Gamma(m+1)$, where $\Gamma(x)$
is the classical Euler gamma-function and note that
$$\Gamma(m+1)(e^{2\pi i m} -1) = 2i e^{\pi i m} \sin \pi m \Gamma(m+1)=  (-1)^{m+1}2\pi i \Gamma^{-1}(-m) $$
because of the reflection property of gamma-function:
$$\Gamma(x) \Gamma(1-x) = \frac{\pi}{\sin \pi x}.$$
We note here a remarkable similarity to Riemann's first proof of the reflection property of the Riemann zeta function
$$\pi^{-\frac{s}{2}}\Gamma\left(\frac{s}{2}\right)\zeta(s) = \pi^{-\frac{1-s}{2}}\Gamma\left(\frac{1-s}{2}\right)\zeta(1-s).$$

\section{Trigonometric case}

It is actually more convenient for us to use the hyperbolic rather than trigonometric functions, but all the results are automatically applied to both cases because of the algebraic nature of BA function.

The trigonometric version of the BA function satisfies the equation
\begin{equation}
\label{Sch2}
\mathcal L^{(n)}_m \Phi^{(n)}_m= -4(\nu, \nu) \Phi^{(n)}_m,
\end{equation}
where
\begin{equation}
\label{CMS}
\mathcal L^{(n)}_m=-\Delta + \sum_{i < j}^n \frac{2m(m+1)}{\sinh^2(x_i-x_j)}
\end{equation}
is the Sutherland operator.
It has the form
\begin{equation}
\label{form-t}
\Phi^{(n)}_m = \frac{P(x, \nu)}{B(x)^m C_m(\nu)}\exp2(\nu, x),
\end{equation}
where $P(x, \nu)$ is a trigonometric polynomial in $x$ and a usual
polynomial in $\nu$ with the leading term $B(x)^m A(\nu)^m$, where
$$B(x) = \prod_{i<j}^n \sinh(x_i-x_j)$$ and
$$C_m(\nu) = \prod_{k=1}^m \prod_{i<j}^n (\nu_i-\nu_j-k).$$ The normalisation constant $C_m(\nu)$
is chosen in such a way that
$$\lim_{x \rightarrow +\infty} \frac{P(x, \nu)}{B(x)^m C_m(\nu)} = 1$$ when $x \rightarrow +\infty$
 in the Weyl chamber $x_1 > x_2 >\dots >x_n$ (see \cite{Ch, Fei}).
In the exponential variables
$$u_i = \exp 2x_i, \, i=1, \dots, n$$ the BA function can be rewritten as
\begin{equation}
\label{phi}
\Phi^{(n)}_m = \frac{Q(u,\nu)}{A(u)^{m}C_m(\nu)} u^{\nu},
\end{equation}
where $u^{\nu} = u_1^{\nu_1} \dots u_n^{\nu_n}$ and $Q(u, \nu)$ is polynomial in $\nu$ with the leading term $A(u)^m A(\nu)^m.$

It is convenient to modify the definition of $A(w,u)$ as follows
$$A^*(w,u) = \prod_{i\leq j} (w_i-u_j) \prod_{i > j} (u_j-w_i)$$
to include some sign factor.

Let $w=(w_1,\dots, w_k),\, u=(u_1, \dots, u_{k+1})$ and the cycle $\sigma$ similarly to the rational case be the product of  circles $|u_i-w_i| = \epsilon,\,  i= 1, \dots, k$ with small positive $\epsilon.$

\begin{thm}
The trigonometric BA functions $\Phi^{(k)}_m(w,\nu_1,\dots,
\nu_{k})$ and $\Phi^{(k+1)}_m(u,\nu_1, \dots, \nu_{k+1})$ are
related by the following iterated residue formula
\begin{equation}
\label{formula1*}
\prod_{i=1}^{k+1}u_i^{\nu_{k+1}} \int_{\sigma}  \frac{A(u)^{m+1}A(w)^{m+1}} {A^*(w,u)^{m+1}} \prod_{i=1}^k w_i^{m-\nu_{k+1} }\Phi^{(k)}_m(w, \nu_1, \dots, \nu_k) dw= C \Phi^{(k+1)}_m(u, \nu_1, \dots, \nu_{k+1})
\end{equation}
where  $$C=(2\pi i)^k \prod_{i=1}^{k} {\nu_i - \nu_{k+1}-1\choose m}$$ and
$${a \choose m} = \frac{a(a-1)\dots(a-m+1)}{m!}.$$
\end{thm}

As before we introduce $\frac{n(n-1)}{2}$ integration variables $t_{i,j}$ with $1 \leq i < j \leq n$ with the convention that $t_{i,n} =u_i, \, i = 1, \dots, n.$
The integrand in the trigonometric case has the following form (cf. \cite{KK})
%\begin{equation}
%\label{omega-t}
$$
\omega^*_m=
\prod_{j=1}^{n-1} A_{j,j+1}^{-m-1}
\prod_{j=1}^{n} A_{j,j}^{2+2m} \prod_{l<j}
\prod_{i \leq j}^{n-1}  t_{i,j}^{\nu_{j} - \nu_{j+1} +m} dt_{i,j},
$$
%\end{equation}
where
$$A_{j,j+1}(t) = \prod_{i \leq j, l\leq  j+1, i\leq l}^{n} (t_{i,j}  - t_{l,j+1}) \prod_{i \leq j, l\leq  j+1, i>l}^{n} (t_{i,j}  - t_{l,j+1})$$
and
$$A_{j,j}(t)= \prod_{1 \leq i < l \leq j \leq n-1} (t_{i,j} - t_{l,j}).$$
The cycle of integration $\Sigma$ as before is the product of circles $|t_{i,j} - u_i| = \epsilon(n- j)$ with $\epsilon$ small enough. Note that in contrast to \cite{KK} the origin is outside of these circles, so we have no problems with the multi-valuedness of the integrand.

\begin{corollary}
The trigonometric BA function (\ref{phi}) can be given as an
iterated residue
\begin{equation}
\label{int-2}
\Phi^{(n)}_m(u,\nu)=  C(n,m, \nu) \prod_{i=1}^{n} u_i^{\nu_1} A(u)^{1+m} \int_{\Sigma} \omega^*_m
\end{equation}
with
$$C(n,m,\nu)^{-1}= (2\pi i)^{\frac{n(n-1)}{2}} \prod_{i<j}^n {\nu_i - \nu_{j}-1\choose m}.$$
\end{corollary}

Similarly to the rational case we have also the following Selberg-type representation of BA functions.

Assume for convenience that the complex numbers $u_1, u_2, \dots, u_{k+1}$  have different arguments and consider the contour $\gamma^*$ when $w_i$ belongs to the segment joining $0$ and $u_i$ for $i=1,\dots, k$. In other words, we assume that $w_i = \tau_i u_i,\, i=1,\dots k$ with real $\tau_i$ between 0 and 1. We assume also that $\nu_i - \nu_{k+1}$ have large positive real parts to guarantee the convergence of the following integral.

\begin{thm}
The BA functions $\Phi^{(k)}_m$ and $\Phi^{(k+1)}_m$ are related by the Selberg-type integral formula
\begin{equation}
\label{formula2*}
\prod_{i=1}^{k+1}z_i^{\nu_{k+1}} \int_{\gamma^*}  \frac{A^*(w,u)^{m}} {A(u)^{m}A(w)^{m}} \prod_{i=1}^k w_i^{-m-1-\nu_{k+1} }\Phi^{(k)}_m(w, \nu_1, \dots, \nu_k) dw= C_3 \Phi^{(k+1)}_m(u, \nu_1, \dots, \nu_{k+1})
\end{equation}
where  $$C_3=(-1)^{km} \prod_{i=1}^{k} \frac{\Gamma(m+1) \Gamma(\nu_i - \nu_{k+1})}{\Gamma(\nu_i-\nu_{k+1}+m+1)}.$$
\end{thm}

Consider
$$\alpha^*_m= \omega^*_{-m-1}=
\prod_{j=1}^{n-1} A_{j,j+1}^{m}
\prod_{j=1}^{n} A_{j,j}^{-2m} \prod_{l<j}
\prod_{i \leq j}^{n-1}  t_{i,j}^{\nu_{j} - \nu_{j+1} -m-1} dt_{i,j}
$$
and choose the contour of integration $\Gamma^*$ such that $t_{i,j} = \tau_{i,j} t_{i,j+1}$ with $\tau_{i,j} \in [0,1].$

\begin{figure}
\begin{picture}(200,200)(0,-30)
%\put(0,0){.}\put(50,0){.}\put(100,0){.}\put(150,0){.}\put(200,0){.}
%\put(0,50){.}\put(50,50){.}\put(100,50){.}\put(150,50){.}\put(200,50){.}
%\put(0,100){.}\put(50,100){.}\put(100,100){.}\put(150,100){.}\put(200,100){.}
%\put(0,150){.}\put(50,150){.}\put(100,150){.}
 \put(15,15){$u_n$}
 \put(25,50){$u_{n-1}$}
 \put(60,92){$u_{n-2}$}
 \put(175,135){\dots}
 \put(100,120){\dots}
 \put(175,95){$u_1$}
 \put(130,-5){$0$}
\scalebox{0.5}{\includegraphics{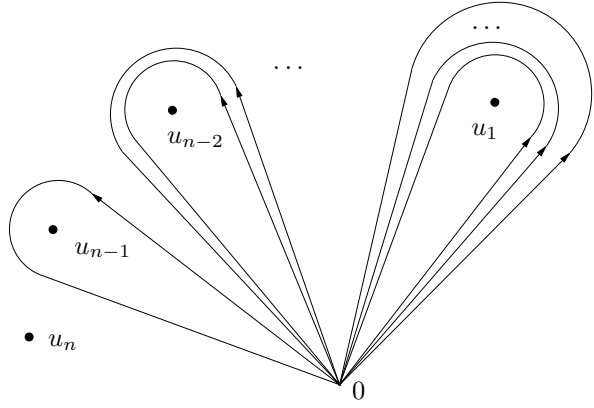}}
\end{picture}
\caption{The Pochhammer contour $\Pi$ in the trigonometric
case}\label{fig2}
\end{figure}

\begin{corollary}
The trigonometric BA function (\ref{phi}) can be given as a
Selberg-type integral
\begin{equation}
\label{int-3}
\Phi^{(n)}_m(u,\nu)=  C_4 \prod_{i=1}^{n} u_i^{\nu_n} A(u)^{1+m} \int_{\Gamma^*} \alpha^*_m
\end{equation}
with
$$C_4 =  (-1)^{m\frac{n(n-1)}{2}}\prod_{i<j}^n\frac{\Gamma(m+1) \Gamma(\nu_i - \nu_{j})}{\Gamma(\nu_i-\nu_{j}+m+1)}.$$
\end{corollary}
 The same calculation as in the rational case shows that these
two integral representations are related by analytic continuation.
The corresponding analogue of the Pochhammer contour is shown in
Figure \ref{fig2}.

%The proof follows directly from the results of Kazarnovski-Krol \cite{KK} and the theory of trigonometric BA function
%\cite{CV, Ch, Fei}.

\section{Deformed case}

Consider now the deformed Calogero--Moser operator \cite{CFV, CFV1}
\begin{equation}
\label{defCM}
L_m^{(n,1)}=-\sum_{i=1}^n \frac{\partial^2}{\partial x_i^2} - m \frac{\partial^2}{\partial y^2}  + \sum_{i < j}^n \frac{2m(m+1)}{(x_i-x_j)^2} + \sum_{i=1}^n\frac{2(m+1)}{(x_i-y)^2},
\end{equation}
corresponding to an additional particle with mass $\frac{1}{m}$
interacting with $n$ Calogero--Moser particles in a special way.
According to \cite{CFV1} for any integer $m \in \mathbb Z$ it has
the eigenfunction (which we will call deformed Baker-Akhiezer
function) of the form
\begin{equation}
\label{defform}
\Psi_m^{(n,1)} = \frac{P(x,y,\lambda, \mu)}{A(x)^{m^*} A(x,y)A(\lambda)^{m^*} A(\lambda,\mu)} \exp((\lambda, x)+ \frac{1}{m} \mu y),
\end{equation}
where $m^* = max(m, -m-1),\, \lambda=(\lambda_1, \dots, \lambda_{n}) \in \mathbb C^n, \, \mu \in \mathbb C,$
$$A(x) = \prod_{i < j}^n (x_i-x_j), \,\, A(x,y) = \prod_{i=1}^n (x_i-y)$$
and $P(x,y, \lambda, \mu)$ is a polynomial in all variables
with the highest degree term $A(x)^{m^*} A(x,y)A(\lambda)^{m^*} A(\lambda,\mu).$

Let $\Psi^{(n)}_m(z_1,\dots, z_n; \lambda_1, \dots, \lambda_n)$ be the BA function (\ref{form}) from the first section. For positive $m$ we assume as before that all $x_i$ have different imaginary parts, real part of $\lambda_i-\mu, \, i=1, \dots, n$ are negative  and choose the contour $\gamma$ of integration $z_i=x_i + \tau_i$ by considering $\tau_i \in \mathbb R_+, \, i=1,\dots, n$.

For negative $m$ we assume simply that $x_i$ are distinct and choose the cycle $\sigma$ as a product of circles $|z_i-x_i|=\epsilon,\, i=1, \dots, n.$

\begin{thm}
The rational deformed Baker-Akhiezer function (\ref{defform}) for positive integer $m$ can be given by the following Selberg-type integral
\begin{equation}
\label{intdef}
\Psi^{(n,1)}_m(x,y, \lambda,\mu) =  \frac{C_5}{A(x)^{m} A(x,y) } \int_{\gamma}\frac{A(z,x)^m A(z,y)}{A(z)^{m}} e^{\mu(\bar x - \bar z+ \frac{y}{m})}\Psi^{(n)}_m(z,\lambda) dz
\end{equation}
with $$C_5 = \frac{\prod_{i=1}^n(\mu - \lambda_i)^{m+1}}{(m!)^n}.$$
For negative $m= -m^*-1,\, m^* \in \mathbb Z_+$ it can be represented as an iterated residue
\begin{equation}
\label{intdefres}
\Psi^{(n,1)}_m =  C_6 \frac{A(x)^{m^*+1}}{A(x,y)} \int_{\sigma}\frac{A(z)^{m^*+1}A(z,y)} {A^*(z,x)^{m^*+1}}  e^{\mu(\bar x - \bar z+ \frac{y}{m})}\Psi^{(n)}_m(z,\lambda) dz,
\end{equation}
where
$$C_6= \left(\frac{m^*!}{2 \pi i}\right)^n \prod_{i=1}^n(\lambda_i-\mu)^{-m^*} .$$
\end{thm}

In the trigonometric case (see \cite{CFV1})
we have the operator
\begin{equation}
\label{defCMtrig}
\mathcal L_m^{(n,1)}=-\sum_{i=1}^n \frac{\partial^2}{\partial x_i^2} - m \frac{\partial^2}{\partial y^2}  + \sum_{i < j}^n \frac{2m(m+1)}{\sinh^2(x_i-x_j)} + \sum_{i=1}^n\frac{2(m+1)}{\sinh^2(x_i-y)},
\end{equation}
and the BA function of the form
\begin{equation}
\label{defformtrig}
\Phi_m^{(n,1)} = \frac{Q(u,v,\nu, \mu)}{A(u)^{m^*} A(u,v)C_{m^*}(\nu) C(\nu,\mu)} u^{\nu} v^{\frac{\mu}{m}},
\end{equation}
where $u_i = e^{2x_i}, i=1,\dots, n, \, v= e^{2y}$, $Q(u,v,\nu,\mu)$ is polynomial in $\nu \in \mathbb C^n,\, \mu \in \mathbb C$ with the leading term $A(u)^{m^*} A(u,v)A(\nu)^{m^*} A(\nu,\mu)$ and
$$C(\nu,\mu) = \prod_{i=1}^n(\nu_i - \mu- \frac{1+m}{2}) =  \prod_{i=1}^n(\nu_i - \mu+ \frac{m^*}{2}).$$
As in the non-deformed case, the form of $C(\nu,\mu)$ is uniquely determined by the property
$$\lim\frac{Q(u,v,\nu, \mu)}{A(u)^{m^*} A(u,v)C_{m^*}(\nu) C(\nu,\mu)} = 1$$ when $(u,v) \rightarrow +\infty$ in the chamber $u_1 > u_2 >\dots > u_n > v.$

Let $\Phi^{(n)}_m(u_1,\dots, u_n, \nu_1, \dots, \nu_n)$ be the non-deformed $n$ particle BA function (\ref{phi}).
For positive $m$ we assume that all $u_i$ have different arguments, all $\lambda_i-\nu$ are real negative  and choose the contour $\gamma^*$ of integration $w_i=\tau_i u_i$ with real $\tau_i \in [0,1].$
For negative $m= -1-m^*$ we assume simply that $u_i$ are distinct and choose the cycle $\sigma^*$ as a product of circles $|w_i-u_i|=\epsilon,\, i=1, \dots, n.$

\begin{thm}
The deformed trigonometric Baker-Akhiezer function $\Phi^{(n,1)}_m(u,v, \nu,\mu) $ for positive integer $m$ can be given by the following Selberg-type integral
\begin{equation}
\label{intdeftrig}
\Phi^{(n,1)}_m = C_7\frac{\prod_{i=1}^n u_i^{\mu+ \frac{1-m}{2}} v^{\frac{\mu}{m}}}{A(u)^m A(u,v)} \int_{\gamma^*}\frac{A(w,u)^m A(w,v)}{ A(w)^{m}} \prod_{i=1}^n w_i^{-\mu - \frac{1+m}{2} }\Phi^{(n)}_m(w,\nu) dw
\end{equation}
with
$$ C_7= \prod_{k=1}^n \frac{\Gamma(m+1) \Gamma(\nu_k-\mu + \frac{1-m}{2})}{\Gamma(\nu_k-\mu+\frac{m+3}{2})} $$
and for negative $m= -m^*-1,\, m^* \in \mathbb Z_+$ as an iterated residue
\begin{equation}
\label{intdefrestrig}
\Phi^{(n,1)}_m = C_8\frac{\prod_{i=1}^n u_i^{\mu+ \frac{m^*+2}{2}} v^{\frac{\mu}{m}} A(u)^{m^*+1}}{A(u,v)} \int_{\gamma^*}\frac{A(w)^{m^*+1} A(w,v)}{A(w,u)^{m^*+1}} \prod_{i=1}^n w_i^{-\mu + \frac{m^*}{2} } \Phi^{(n)}_m(w,\nu) dw
\end{equation}
where
$$C_8^{-1}= (2 \pi i)^n \prod_{i=1}^{n} {\nu_i-\mu + \frac{m^*}{2} \choose m^*}.$$
\end{thm}

\section{Proofs: main identity}

Let
\begin{eqnarray}
\label{anm} L_m^{k,l}(x,y)& = &-\left (\frac{\partial^2}{{\partial
x_{1}}^2}+\dots +\frac{\partial^2}{{\partial x_{k}}^2}\right)-
m\left(\frac{\partial^2}{{\partial y_{1}}^2}+\dots
+\frac{\partial^2}{{\partial y_{l}}^2}\right)
+\sum_{i<j}^{k}\frac{2m(m+1)}{\sinh^2(x_{i}-x_{j})} \nonumber \\ &
&+\sum_{i<j}^{l}\frac{2(m^{-1}+1)}{\sinh^2(y_{i}-y_{j})}
+\sum_{i=1}^{l}\sum_{j=1}^{k}\frac{2(m+1)}{\sinh^2(x_{i}-y_{j})}
\end{eqnarray}
and
\begin{eqnarray}
\label{anmrat} L_m^{k,l}(x,y)& = &-\left (\frac{\partial^2}{{\partial
x_{1}}^2}+\dots +\frac{\partial^2}{{\partial x_{k}}^2}\right)-
m\left(\frac{\partial^2}{{\partial y_{1}}^2}+\dots
+\frac{\partial^2}{{\partial y_{l}}^2}\right)
+\sum_{i<j}^{k}\frac{2m(m+1)}{(x_{i}-x_{j})^2} \nonumber \\ &
&+\sum_{i<j}^{l}\frac{2(m^{-1}+1)}{(y_{i}-y_{j})^2}
+\sum_{i=1}^{l}\sum_{j=1}^{k}\frac{2(m+1)}{(x_{i}-y_{j})^2},
\end{eqnarray}
be respectively trigonometric and rational deformed %%%
Calogero--Moser--Sutherland (CMS) operators in $x,y$ variables \cite{SV1}. Let
$L_m^{p,q}$ be a similar operator in variables $z_1, \dots, z_p,
w_1, \dots, w_q.$
Let $A(u,v)$ be given by (\ref{A}),
$$B(u,v)= \prod_{i=1}^k \prod_{j=1}^l \sinh(u_i-v_j)$$
 and $\bar x=x_1 + \dots +x_k$ as before.

The key observation \footnote{As we have learnt from Martin Hallnas a similar result can be extracted from his paper with Edwin Langmann \cite{HL}.}  comes from the following result (cf. Langmann \cite{Lan}).

\begin{thm}
The following identity holds for the trigonometric %%%
deformed CMS operators:
\begin{equation}
\label{ident}
L_m^{k,l} K = L_m^{p,q} K + C_0 K,
\end{equation}
where $$K(x,y; z,w) =  \frac{B(x,z)^m B(y,z) B(x,w) B(y,w)^{1/m}}{B(x)^{m} B(x,y) B(y)^{1/m} B(z)^{m}B(z, w) B(w)^{1/m}}e^{\mu(\bar x - \bar z+\frac{1}{m}(\bar y -\bar w))}$$
and
$$C_0= \frac{1}{4} m^2 [(k-p+\frac{1}{m}(l-q))^3 - (k-p+\frac{1}{m^3}(l-q))] +\mu^2(p+\frac{q}{m}-k-\frac{l}{m}).$$

In the rational case we have the same identity (\ref{ident}) for
$$K(x,y; z,w)  = \frac{A(x,z)^m A(y,z) A(x,w) A(y,w)^{1/m}}{A(x)^{m} A(x,y) A(y)^{1/m} A(z)^{m}A(z, w) A(w)^{1/m}}e^{\mu(\bar x - \bar z+\frac{1}{m}(\bar y -\bar w))}$$
and $C_0= \mu^2(p+\frac{q}{m}-k-\frac{l}{m}).$
\end{thm}

The identity (\ref{ident}) with $l=q=0$ goes back to Stanley and Macdonald \cite{Stanley,Ma}. In the case $q=0$ and arbitrary $l$ it appeared in \cite{SV2} (see part (iii) in Lemma 3). To prove it in the general case we borrow the idea from the work of Langmann \cite{Lan}.
It is based on the following result by Sen \cite{Sen} (see also \cite{KO}).

Consider the following generalised CMS operator describing the interacting particles of different masses on a line:
\begin{equation}
\label{gensen}
H = -\sum_{j=1}^N \frac{1}{m_j} \frac{\partial^2}{\partial x_j^2} + \sum_{j<k} \frac{\gamma_{jk}}{\sinh^2(x_j - x_k)}.
\end{equation}

\begin{thm} [Sen \cite{Sen}]
For the coupling constants of the special form
\begin{equation}
\label{sform}
\gamma_{ij} = (m_i + m_j)\beta (m_i m_j \beta -1),
\end{equation}
where $\beta$ is an arbitrary parameter,
the operator (\ref{gensen}) has the following eigenfunction
\begin{equation}
\label{seigen}
\Phi_0 = \prod_{i<j} \sinh^{\beta m_i m_j}(x_i-x_j):
\end{equation}
$$H \Phi_0 = E_0 \Phi_0$$
with the eigenvalue
\begin{equation}
\label{seigenv}
E_0 = -\frac{\beta^2}{3}((\sum_{j=1}^N m_j)^3 - \sum_{j=1}^N m_j^3).
\end{equation}
\end{thm}

Now we note that $\gamma_{ij} = 0$ if $m_j=-m_i$ or $m_j = (m_i\beta)^{-1}.$
Choosing $$m_1=\dots = m_k=1,\, m_{k+1} = \dots = m_{k+l} = m^{-1},$$
$$ m_{k+l+1} = \dots = m_{k+l+p} = -1, \, m_{k+l+p+1} = \dots = m_{k+l+p+q} = -m^{-1},$$
where $m = -\beta,$ we see that the operator $H$ reduces to the difference
$$H= L_m^{k,l} - L_m^{p,q}$$
of two decoupled deformed CMS operators and the relation $H \Phi_0 = E_0 \Phi_0$
implies the identity (\ref{ident}).

We note that this decoupling can be used to characterise the deformed CMS operators among
all generalised CMS operators with different masses (\ref{gensen}). It does not imply though the quantum integrability, which had to be proven by other means (see \cite{SV1, SV2}).

The identity (\ref{ident}) suggests that the function $K(x,y,z,w)$ could be used as the kernel of the integral representation transforming the eigenfunctions of one of the deformed CMS operators to another, although to make this precise could be a non-trivial task (see e.g. \cite{R}, where a
similar problem is discussed).

In particular, choosing in the rational case $k=p+1, l=q=0$ we come to the integral of the type (\ref{formula1*}). Choosing a suitable cycle (contour) of integration we come to the formula for the BA function. For example, in Theorem \ref{th1} the cycle $\sigma$ is chosen in such a way that each integration has only one non-zero residue to compute, which guarantee that the result will be of the required form.

The rest of the proofs follows from the results of the papers \cite{VSCh, CFV, Ch, Fei} on the Baker-Akhiezer functions.

\section{Examples}

In the simplest case $n=2$ the rational BA function is known to have the form (see e.g. \cite{CV}):
\begin{equation}
\label{psi2}
\Psi^{(2)}_m =(\lambda_1 - \lambda_2)^{-m}(D_{12}-\frac{2m}{x_1-x_2})(D_{12}-\frac{2(m-1)}{x_1-x_2})\dots (D_{12}-\frac{2}{x_1-x_2})\exp(\lambda_1 x_1 + \lambda_2 x_2),
\end{equation}
where $$D_{12} = \frac{\partial}{\partial x_1}- \frac{\partial}{\partial x_2}.$$

We have two different representations for it. The first one is as a residue
\begin{equation}
\label{psi2it}
\Psi^{(2)}_m = \frac{m! (x_1-x_2)^{m+1}}{(\lambda_1-\lambda_2)^m}e^{\lambda_2(x_1+x_2)} Res_{z=x_1} \frac{e^{(\lambda_1-\lambda_2)z}}{(z-x_1)^{m+1}(z-x_2)^{m+1}},
\end{equation}
the second one is the integral
\begin{equation}
\label{psi2int}
\Psi^{(2)}_m = \frac{(\lambda_2-\lambda_1)^{m+1}}{m! (x_1-x_2)^m}e^{\lambda_2(x_1+x_2)} \int_{x_1}^{+\infty} (z-x_1)^{m}(z-x_2)^{m} e^{(\lambda_1-\lambda_2)z} dz,
\end{equation}
which in this case can be effectively computed using the $\Gamma$-integral
$$\Gamma(a) = \int_{0}^{+\infty} z^{a-1} e^{-z}dz = (a-1)!$$ for positive integer $a$.

For $n=3$ the corresponding BA function $\Psi^{(3)}_m(x_1,x_2,x_3, \lambda_1, \lambda_2, \lambda_3)$ can be written respectively as follows:
\begin{equation}
\label{psi3it}
\Psi^{(3)}_m = C Res_{z_1=x_1} Res_{z_2 = x_2} Res_{w=z_1} \frac{(z_1 - z_2)^{2m+2}e^{(\lambda_2-\lambda_3)\bar z} e^{(\lambda_1 - \lambda_2)w}}{\prod_{i=1}^2(w- z_i)^{m+1}\prod_{i=1}^{2}\prod_{j=1}^3(z_i-x_j)^{m+1}}
\end{equation}
where $$C= \frac{(m!)^3 \prod_{i<j}^3 (x_i-x_j)^{m+1}}{ \prod_{i<j}^3 (\lambda_i-\lambda_j)^{m}}e^{\lambda_3\bar x}$$
\begin{equation}
\label{psi3int}
\Psi^{(3)}_m = D \int_{x_1}^{+\infty} \int_{x_2}^{+\infty}\int_{z_1}^{+\infty}\frac{\prod_{i=1}^2(w- z_i)^{m}\prod_{i=1}^{2}\prod_{j=1}^3(z_i-x_j)^{m} e^{(\lambda_2-\lambda_3)\bar z} e^{(\lambda_1 - \lambda_2)w}}{ (z_1 - z_2)^{2m}} dz_1 dz_2 dw,
\end{equation}
with
$$D= (-1)^{m+1}\frac{ \prod_{i<j}^3 (\lambda_i-\lambda_j)^{m+1}}{(m!)^3 \prod_{i<j}^3 (x_i-x_j)^{m}}e^{\lambda_3\bar x}$$
where as before $\bar x = x_1+x_2+x_3, \bar z =  z_1+z_2.$

One can interpret these formulas either as a new way of representing of BA function, or as an explicit evaluation of the Selberg-type integral (\ref{psi3int}) in terms of the BA function, which can be computed by other methods as well (see \cite{CV, VSCh, CFV}). The same of course is true for general $n$ and in the deformed case.

\section{Concluding remarks}

It would be interesting to explore the possibilities of choosing different cycles to produce the integral formulas for the super Jack polynomials \cite{SV2}. For the usual Jack polynomials such integral formulas were obtained in \cite{Awata, OO, KMS, Lan}.

Our approach can be also naturally extended to the (deformed) $BC_n$-case and related theory of (super) Jacobi polynomials \cite{SV3}.
We will discuss this in more detail elsewhere.

\section*{Acknowledgements} We would like to thank O. Chalykh, M. Hallnas, S. Ruijsenaars, A.N. Sergeev and especially M. Feigin for very useful discussions and comments.
APV is grateful to Forschungsinstitut f\"ur Mathematik, ETH, Zurich for the hospitality during the spring 2008 when the work has been done.

 This work has been partially supported by the
European Union through the FP6 Marie Curie RTN ENIGMA (Contract
number MRTN-CT-2004-5652) and ESF programme MISGAM and by the EPSRC (grant EP/E004008/1).

\end{document}